\documentclass[superscriptaddress,aps,prl,twocolumn]{revtex4-1}
\usepackage{graphicx}
\usepackage{textcomp}
\usepackage{amssymb}
\usepackage{amsmath}
\usepackage[utf8]{inputenc}

\usepackage{chemformula} 
\usepackage[T1]{fontenc}

\usepackage{amsfonts}
\usepackage{mathbbol}
\usepackage{mathptmx}

\newcommand{\dd}[0]{\mathrm{d}}
\begin{document}
\title{Wrinkling instability in 3D active nematics}
\author{Tobias Str{\"u}bing}
\affiliation{Max Planck Institute for Dynamics and Self-Organization (MPIDS), 37077 G\"ottingen, Germany}
\author{Amir Khosravanizadeh}
\affiliation{Max Planck Institute for Dynamics and Self-Organization (MPIDS), 37077 G\"ottingen, Germany}
\affiliation{Department of Physics, Institute for Advanced Studies in Basic Sciences, Zanjan 45137-66731, Iran}
\author{Andrej Vilfan} \affiliation{Max Planck Institute for Dynamics and
  Self-Organization (MPIDS), 37077 G\"ottingen,
  Germany}
\affiliation{Jo\v{z}ef Stefan Institute, 1000 Ljubljana, Slovenia}
\author{Eberhard Bodenschatz} \affiliation{Max Planck Institute for
  Dynamics and Self-Organization (MPIDS), 37077 G\"ottingen,
  Germany}
\affiliation{Institute for Dynamics of Complex Systems,
  Georg-August-University G{\"o}ttingen, 37073 G{\"o}ttingen, Germany}
\affiliation{ Laboratory of Atomic and Solid-State Physics,
  Cornell University, Ithaca, NY 14853, United States} 
\author{Ramin Golestanian} \affiliation{Max Planck Institute for Dynamics and Self-Organization (MPIDS), 37077 G\"ottingen,
  Germany}
\affiliation{Rudolf Peierls Centre for Theoretical Physics,
  University of Oxford, Oxford OX1 3PU, United Kingdom}
\author{Isabella Guido} \email{isabella.guido@ds.mpg.de}  \affiliation{Max Planck
  Institute for Dynamics and Self-Organization (MPIDS), 37077
  G\"ottingen, Germany}
\begin{abstract}
  In nature interactions between biopolymers and motor proteins give
  rise to biologically essential emergent behaviours. Besides
  cytoskeleton mechanics, active nematics arise from such
  interactions.  Here we present a study on 3D active nematics made of
  microtubules, kinesin motors and depleting agent. It shows a rich
  behaviour evolving from a nematically ordered space-filling
  distribution of microtubule bundles toward a flattened and
  contracted 2D ribbon that undergoes a wrinkling instability and
  subsequently transitions into a 3D active turbulent state. The
  wrinkle wavelength is independent of the ATP concentration and our
  theoretical model describes its relation with the appearance
  time. We compare the experimental results with a numerical
  simulation that confirms the key role of kinesin motors in
  cross-linking and sliding the microtubules. Our results on the
  active contraction of the network and the independence of wrinkle
  wavelength on ATP concentration are important steps forward for the
  understanding of these 3D systems.
\end{abstract}
\maketitle
\section{Introduction}
Active matter is a state of matter where large-scale dynamical
structures emerge from the interaction of individual active components
each driven by their own internal energy source.  A well investigated
class of such systems comprises active nematics
\cite{Marchetti2013,Sagues_review} where individual elongated
components self-organize into a dynamic state with spatio-temporal
chaos with topological defects -- a behaviour known as active
turbulence \cite{Kessler,Zhou1265,Zhang13626, Sano,Chate2017,
  Uchida,Saw, Sano2}.  Several experimental and theoretical model
systems have been developed to study networks of cytoskeletal
filaments and motor proteins fueled by ATP
\cite{BauschPNAS,Sanchez,Alper}.  A remarkable example is the emergent
behaviour in 2D observed in mixtures of microtubules and kinesin
motors arranged in bundles by a depletion agent pioneered by Sanchez
et al. \cite{Sanchez}. Internally generated spatio-temporally chaotic
flows at the millimetre scale were observed that persisted as long as
ATP was available. The 2D networks exhibit a steady state with
permanent flow at large scales, while the dynamics is driven by the
buckling and elongation of the microtubule bundles due to the activity
of the motors on the nanometer scale \cite{Voituriez}. At high
concentrations these networks show a transition to a locally ordered
nematic liquid-crystalline state with topological defects and
spatio-temporal dynamics
\cite{GiomiPRL,Yeomans2013,Thampi2014,Giomi2014,Yeomans2014,Yeomans_J,Sagues_Marchetti}
that can be varied by confinement such as structured liquid interfaces \cite{Sagues1}, hard wall circular boundaries  \cite{Dogic2019}, curved surfaces of deformable spherical vesicles \cite{Keber1135}, toroidal droplets \cite{Fernandez}as well as by radial alignment \cite{Sagues2019}. These systems share the experimental approach to be condensed at an oil-water interface.\\
Whereas systems investigated in two-dimensions were the focus of
copious studies, the prospect of developing three-dimensional active
matter systems was shown only by a few examples.  These 3D experiments
included microtubule-based active gels confined in toroids
\cite{Wueaal1979}, and sedimented cell-sized
droplets \cite{Suzuki201616001}.\\
In the present study we show experimentally that a three-dimensional
suspension of microtubules and kinesin-1 motors self-organizes into an
active nematic ribbon that for a range of parameters wrinkles
periodically in the third-dimension due to its longitudinal extension
before transitioning to a state of active turbulence. Similar, but
different, results of a parallel study have been recently published
\cite{Senoussi22464}. However, the contraction in the mentioned system
is passive and due to depletion agent, while our network actively
contracts due to the motor action. A related folding phenomenon has
been observed also in thin acto-myosin networks \cite{Bernheim} whose
behaviour was generated by contracting forces and not by extensile
ones as in the present work. Additionally we understand the behaviour
of our system with stochastic simulations and a quantitative theory
that describes the wrinkling instability.

\section{Results and discussion}
\begin{figure*}
	\centering
	\includegraphics[width=\linewidth]{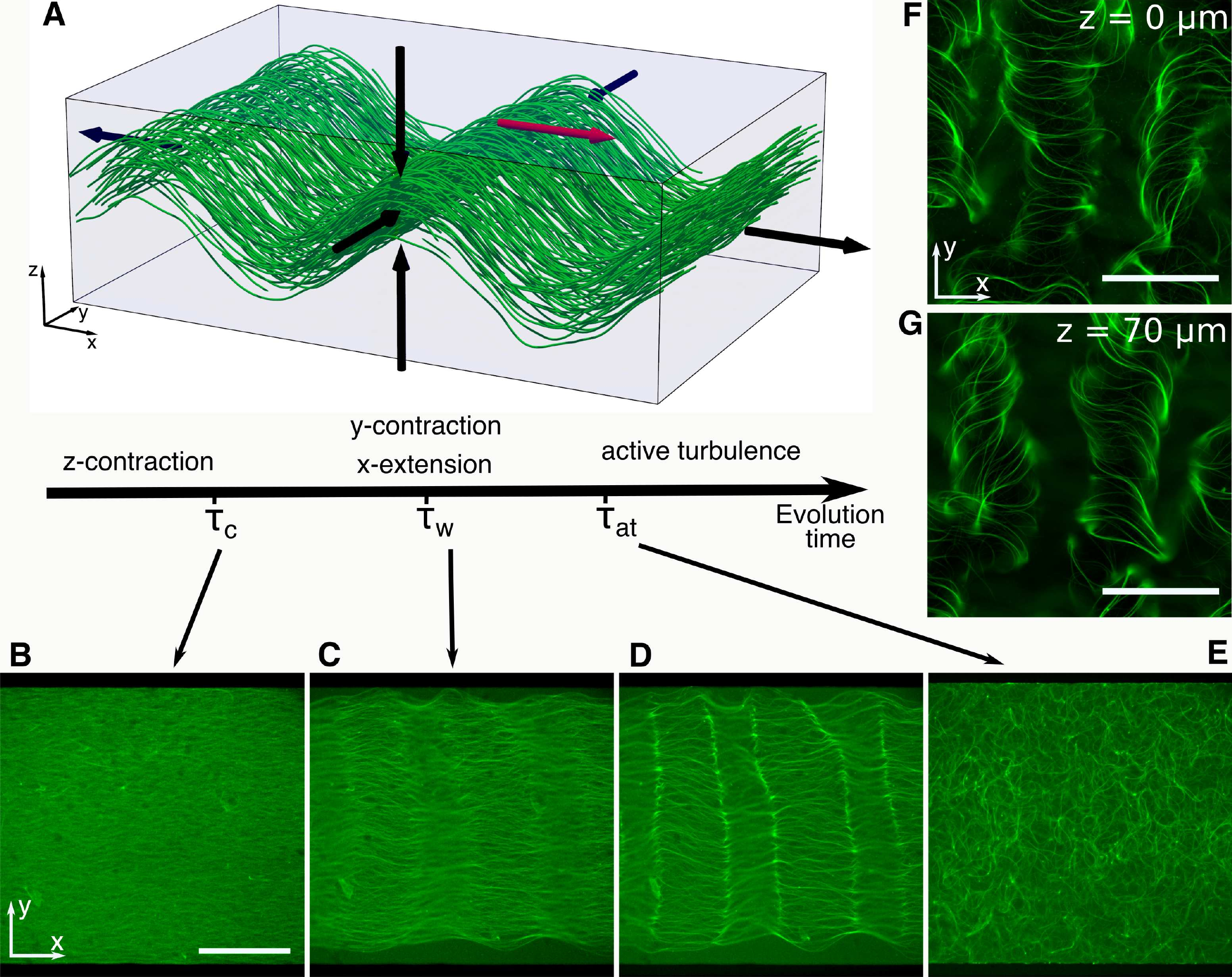}
	\caption{\textbf{3D wrinkling formation}. (\textbf{A}) Schematic representation of the wrinkling process as a result of contractile and extensile forces that emerge at different times: the 3D active nematic system evolves to exhibit contraction into a sheet, a wrinkling instability simultaneously as a lateral contraction, and a final transition to active turbulence with spatio-temporal chaos. The red arrow indicates the nematic director before the onset of wrinkling instability. The black arrows indicate the elastic deformations induced by the active stress along different directions. (\textbf{B--E}) Time evolution of 3D active nematics through the different stages. The nematic order along the $x$-direction is visible. Scale bar: 500 $\mu$m. (\textbf{F--G}) Pattern visualization at two different heights. The 3D structure can be seen by the wave crests in focus and the adjacent ones out-of-focus when the focal plane is set to $z = 0$ $\mu$m. At $z = 70$ $\mu$m the situation is reversed. The experiment was conducted at 2mM ATP. Scale bar: 250 $\mu$m.}
	\label{Fig:Figure1}
\end{figure*}

\paragraph*{\textbf{Experimental results.}}
We  experimentally investigate the dynamics of an active nematic, confined in a long rectangular channel of size 30 mm $\times$ 1.5 mm $\times$ 100 $\mu$m. The active nematic consists of an ionic aqueous solution of microtubules, multi-headed kinesin-1 motors, and poly(ethylene glycol) (PEG) as depletion agent \cite{Sanchez}. We observe that when the system is initially prepared with nematic order \cite{Prost2015} along the long axis of the channel, it exhibits a sequence of transitions before developing into the active turbulent phase with spatio-temporal chaos. This is summarized in Fig.~\ref{Fig:Figure1}.

\begin{figure}[t]
	\centering
	\includegraphics[width=\linewidth]{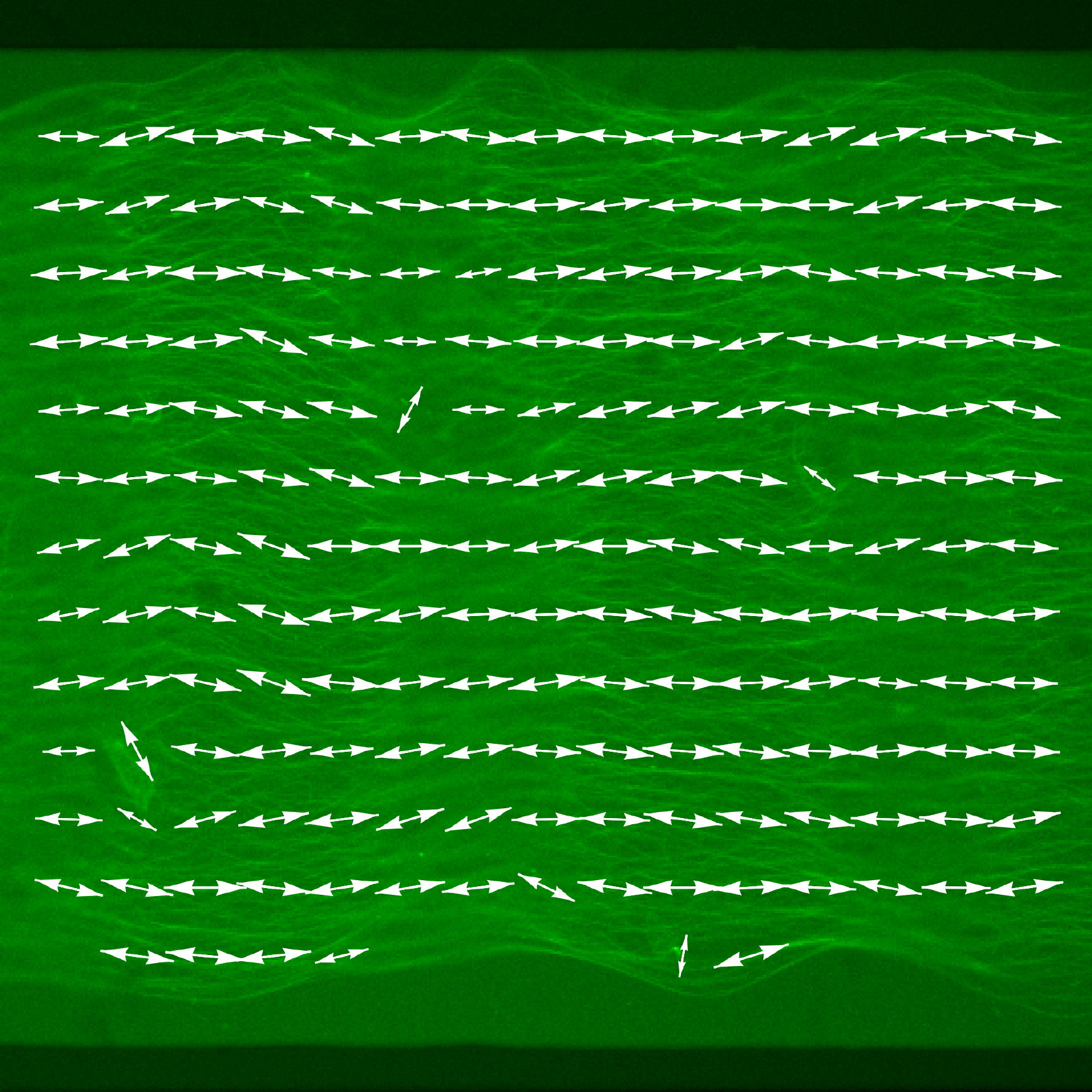}
	\caption{\textbf{Reconstruction of the nematic director field}. This confirms the nematic order of the microtubule bundles in the samples. The image is from Fig.~\ref{Fig:Figure1}C. For details see Supplementary Methods.}
	\label{Fig:Fig2}
\end{figure}

In particular, the system makes a transition from a space-filling 3D active nematic to a narrow ribbon extending over the length of the channel and located in the mid-plane (away from the boundaries). This initial contraction along the $z$-direction, which occurs over a time scale of $\tau_{\rm c}$, is followed by a wrinkling instability at time scale $\tau_{\rm w}$, which takes the system back to being 3D. The long-lived state of spatially periodic wrinkling that is oriented along the long axis undergoes another transition to an active turbulent state where the active nematic fills the entire channel after a long time $\tau_{\rm at}$. This remarkable sequence demonstrates 3D active nematic dynamics and instabilities similar to those observed in a recent study \cite{Senoussi22464}. Conversely, no comparison can be made with 2D systems investigated so far \cite{Sagues2019} due to different spatial distribution of that system and its interaction with substrates.\\
Inside the ribbon nematically ordered filament bundles are clearly visible. To confirm this bundles organization we quantify the nematic order in the sample by using correlation analysis and reconstruct the nematic director field as can be seen in Fig.~\ref{Fig:Fig2}.
The first mode of evolution of the active nematic is contraction perpendicular to the nematic director. After contraction, extensile stress builds up along the nematic director \cite{Marchetti2013} (Fig.~\ref{Fig:Figure1}A). Here, contractile and extensile stresses are internal stresses that would lead to contraction and extension, respectively, in the absence of constraints. In the channel, long-range expansion along the whole channel length is not possible and the extensile stress can lead to a spatially periodic wrinkling of the ribbon (Fig.~\ref{Fig:Figure1}B-E). 
No in-plane bending (horizontal deformation of the director field, akin to deformations in 2D active nematics) is observed at this stage. We expect that out-of-plane bending is preferred to in-plane deformations because it is opposed mainly by the stiffness of the microtubules whereas the in-plane bending additionally requires shearing of the cross-linked sheet.
By acquiring consecutive areas along the $x$-axis we showed that the 3D instabilities are not local but rather emerging in the whole network over centimeters (Fig.~S3). As the wrinkles grow, they reach the channel walls and start folding over. At this point, the sheet-like structure disappears and the individual bundles continue to extend in different directions in space. 
After the time $\tau_{\rm at}$, the system reaches a dynamic state of active turbulence (Fig.~\ref{Fig:Figure1}E).
The characteristics of the material needed to form a 3D active nematic differ significantly from those in the previously studied 2D situations \cite{Sanchez, Sagues2019}. Those studies typically use short microtubules with lengths around 1$\mu$m, higher depletant concentration, an oil-water interface and a radial arrangement. Nematic ordering and active stress buildup in our 3D systems have a relatively low volume fraction of microtubules ($~0.001$), require longer filaments. The average microtubule length in our experiment was 19 $\mu$m $\pm$  10 $\mu$m (Fig.~S2). We verified the importance of length by shearing the filaments to much shorter lengths, in which case no pattern was observed.\medskip

\paragraph*{\textbf{Contraction characterization.}}
\begin{figure}
	\centering
	\includegraphics[width=\linewidth]{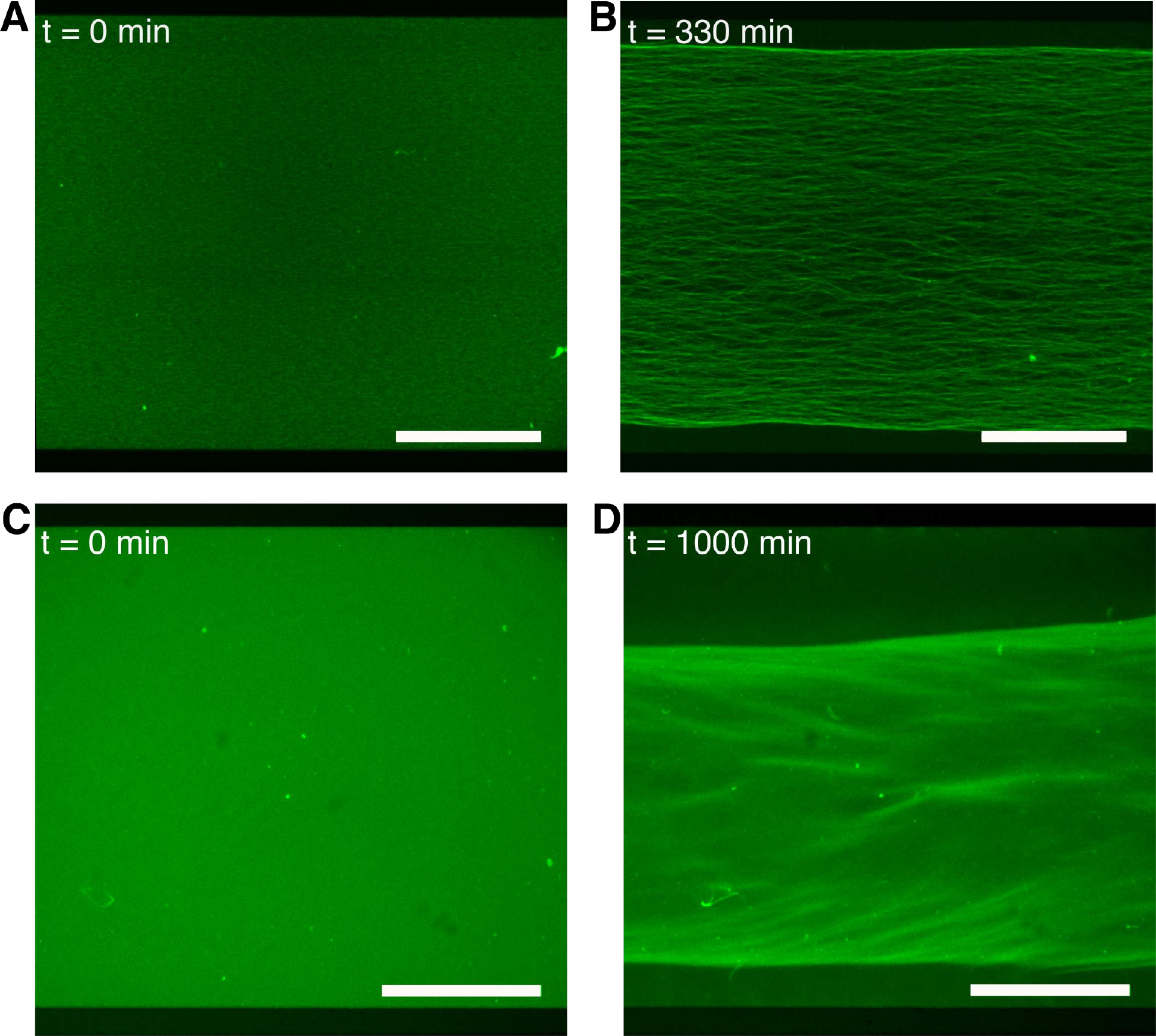}
	\caption{\textbf{Contraction with and without depletion agent PEG}. (\textbf{A--B}). Lateral contraction over time of a network made of microtubule and PEG mixture without addition of motor proteins. The system passively contracts due to the depletion effect of PEG. (\textbf{C--D}). Lateral contraction and longitudinal expansion of a microtubules-motor proteins network over time without addition of PEG. The contracting effect due to the motors proves that the network exhibits an active contraction contribution. The experiment was conducted at 2mM ATP concentration. Scale bar: 500 $\mu$m}
	\label{Fig:with_woPEG}
\end{figure}
The evolution of the system from 3D to 2D includes contraction along both y- and z-directions. The average lateral ($y$) contraction was 169 $\pm$ 8 $\mu$m or 11.2$\%$. Vertically ($z$-direction),  the ribbon thickness  reaches 30 $\mu$m $\pm$ 2 $\mu$m, corresponding to a shrinkage by 70$\%$.
The non-uniform contraction likely reflects the dimensions of the channel. Lateral contraction involves larger absolute displacements and is therefore opposed by stronger drag forces. Similar non-uniform contraction is also seen in the simulation (Fig.~\ref{Fig:Fig5}H-I). 
Whereas Senoussi et al. \cite{Senoussi22464} attribute the contraction to the effect of the depletion agent alone (2$\%$ Pluronic in that study), we show that in our system it results from the active component due to cross-linking activity of kinesin motors combined with the passive bundling effect of PEG. 
By carrying out the experiment without including motors to the filament mixture we could observe the bundling of microtubules due to PEG alone, their nematic orientation along the longitudinal axis of the channel and the y-contraction ( Fig.~\ref{Fig:with_woPEG}A-B). The network contracts by 7.6$\%$ and reaches its final, stable state after $\sim$330 min. No instabilities are present. The final thickness of ca. 40-45 $\mu$m corresponds to a contraction of 57$\%$.  
These contractions can be explained by the entropic mechanism of the depletion interaction. The non-adsorbing polymer (PEG) bundles the microtubules in order to minimize the excluded volume \cite{Braun}. The dimension of the resulting bundles  depends on the molecular weight and radius of gyration of PEG (in our case 20KD and 7 nm, respectively).\newline
In order to show the active contribution to the contraction, we repeated the experiment with motor proteins alone and without PEG. Interestingly, the system actively contracts as well as in the case with depletion agent alone (Fig.~\ref{Fig:with_woPEG}C-D) and reaches a maximum shrinkage of 36$\%$, much stronger then the passive contraction through depletion forces. However, the time scale for shrinkage was 3 times slower than in the passive case. Although in this case the microtubules are ordered in a parallel fashion neither thick bundles (Fig.~S5) nor instabilities are visible. We explain the absence of instability by the missing bundling effect of PEG, which normally reduces the relative distance between microtubules to nanometric gap comparable to the dimension of motor clusters. This represents an optimal value for the continuous binding of kinesin motors and therefore for their efficacy.
The development of the extensile stress that is needed for the wrinkling to occur also requires the combined action of the multiheaded kinesins and the depletion forces. Again the PEG-induced force that holds the filaments together reduces the distance between the microtubules within the bundles, enhances the binding of the motors to the polymers and gives the system the possibility to contract. As a positive feedback reduced distance brings to further contraction given by the cross-linking of the motors. Under this configuration, the system reaches the ribbon configuration and the motors extend the polymers randomly oriented within the ribbon. The extensile stress can be explained by an asymmetric distribution of kinesin motors along microtubules. Because kinesins move towards the microtubule plus ends and then exert a minus-directed force on them, there will be an excess of compressive loads on the microtubules. These, in turn, lead to an extensile stress in the sheet (see Sect.\ Simulation). The ability of the motors to buckle the network was proven by reducing 10-fold the motors concentration. In this case the system is able to contract but the force of the motors is not enough for buckling the system (Fig.~\ref{Fig:Fig4}E).\medskip

\paragraph*{\textbf{Wavelength analysis.}}
In our experiments at onset of the wrinkling instability the observed wavelength $\lambda$ ranged from 200 $\mu$m to 500 $\mu$m for constant ATP concentration (2 mM) and the time until the onset of instability ($\tau_{\rm w}$) varied between 10 minutes and 2 hours (Fig.~\ref{Fig:Fig4}D). We find that the wavelength of the pattern and the time scale of its formation are correlated -- wrinkling instabilities that appear at shorter times tend to have shorter wavelengths than those forming later.  
\begin{figure*}
	\centering
	\includegraphics[width=\linewidth]{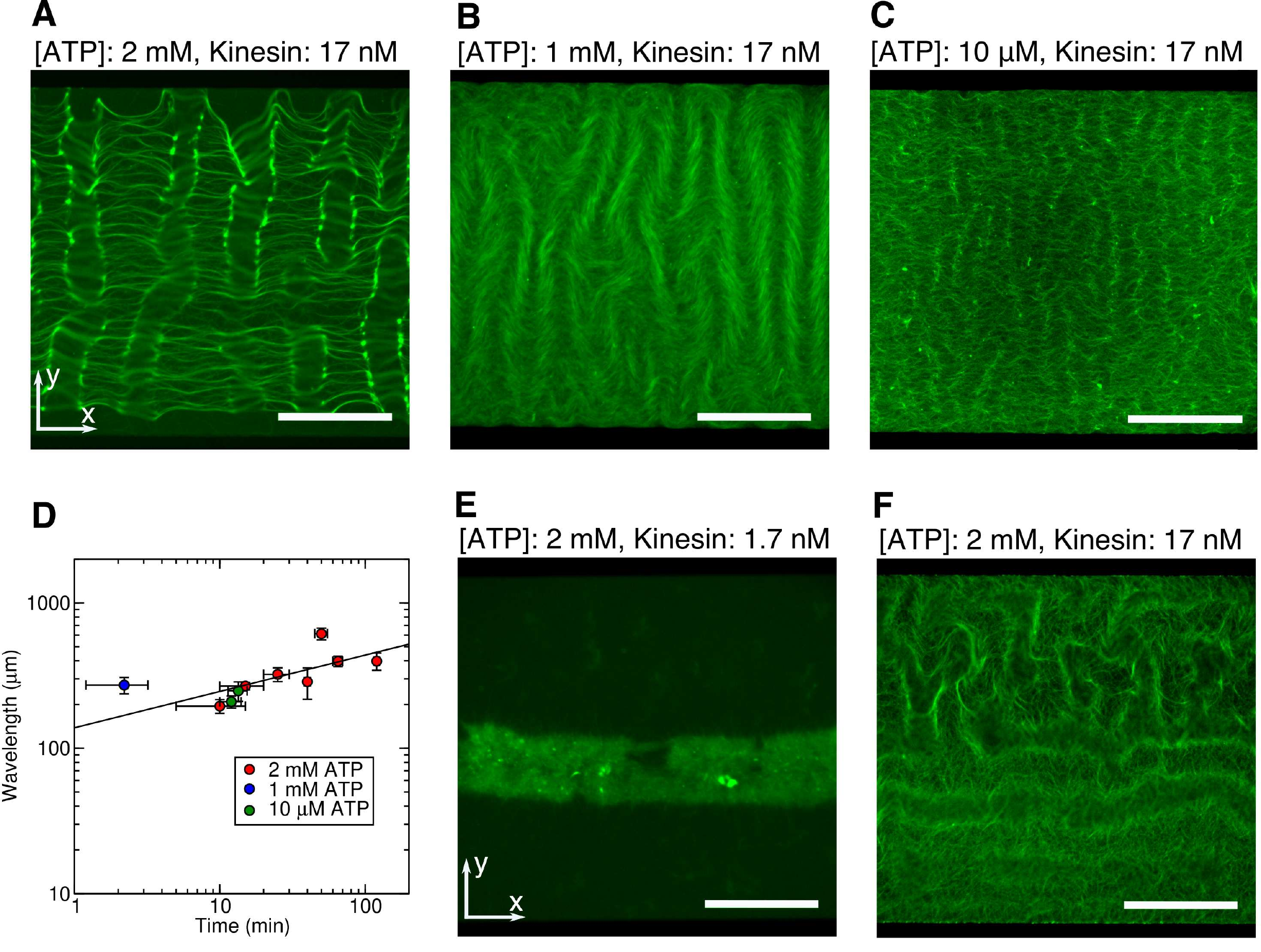}
	\caption{\textbf{Different ATP and motors concentration}. (\textbf{A--C}) Wrinkle formation at different ATP concentrations: 2 mM; 1 mM; limiting concentration 10 $\mu$M. The concentration of motor proteins is 17 nM and constant in the three experiments. (\textbf{D}) The wrinkling wavelength, plotted against the appearance time. The continuous line shows the dependence $\lambda\propto \tau_{\rm w}^{1/4}$, predicted from the model for samples that differ only in the amount of active stress. (\textbf{E}) When the kinesin concentration is reduced by 10-fold, a strong lateral collapse is observed. (\textbf{F}) In some samples, pattern formation can be observed in both $x$-direction and $y$-direction. Scale bars: 500 $\mu$m}
	\label{Fig:Fig4}
\end{figure*}
To further investigate this correlation, we varied the ATP concentration from 2 mM, at which the motors will be reaction-limited, to 10 $\mu$M that is in the diffusion-limited regime.  At 1 mM ATP concentration the wavelength of the pattern was 272 $\mu$m $\pm$ 36 $\mu$m and the wrinkling set in after few minutes; at 10 $\mu$M ATP the wavelength was 249 $\mu$m $\pm$ 38 $\mu$m and the onset of wrinkling occurred at around 13 minutes.
Contrary to what one might expect, no discernible effect on the wavelength of the wrinkling instability could be observed. In other studies 2D system shows a faster response and a shorter selected wavelength with increasing ATP concentration \cite{Sagues2019}. However, due to the differences between the two setups (2D system, high depletant concnetration, confinement of the system to an oil-water interface) no direct comparison can be made. Remarkably in our case the instability does not change with ATP concentration as the time for wrinkling and its wavelength at 10 $\mu$M fall into the range observed at 2mM ATP. This is an interesting result as the variation of the  characteristic wrinkling formation times cannot be understood through the biochemical process linking velocity and force to ATP hydrolysis, as is the case in classical 2D active nematics \cite{Dogic_SM2019}. It was not observed in the other 3D studies as those systems were investigated at constant ATP concentration at reaction-limited regime \cite{Senoussi22464, Bernheim}.
These observations suggest that the active stress is determined by the stall force of kinesin motors, which is independent of the ATP concentration, and by the cross-linking nature of the motors.
To test the importance of the motor concentration, we decreased the kinesin concentration by 10-fold (to 1.7 nM) and indeed did not observe the instability (Fig.~\ref{Fig:Fig4}E). However, the network contracted continuously for around 48 hours resembling the contraction of the stabilized microtubule network in \textit{Xenopus} oocyte extracts \cite{NeedlemanXen}. This result proves that the active force exerted by a lower amount of motors is able to contract the network through the capacity of kinesin to cross-link the filaments, but is not strong enough to trigger the instability. \medskip

\paragraph*{\textbf{Theory.}}
We can develop a theoretical understanding of the observed phenomena based on the combination of the contraction and the mechanical activity inside the resulting ribbon. The director field is mostly aligned in the direction of the channel, $\hat n\simeq \hat e_x$. For microtubules and kinesin motors, we expect an extensile stress along the director field. This stress is accompanied by contractile stresses across the channel (directions $\hat e_y$ and $\hat e_z$). 
After the contractile stress in the $z$-direction leads to the formation of the ribbon, we can describe the system as a two dimensional active nematic sheet that can explore out-of-plane bending deformations. For this effective description, the nematic order is described by the symmetric traceless tensor $\mathbb{Q}=(\hat n \otimes \hat n - \frac 12 \mathbb{I}) S$, with the order parameter $S$, which has the value 1 for perfect nematic alignment and 0 for an isotropic state. The active stress that is caused by the motors can be written as \cite{Simha-Ramaswamy,Prost2015,Sagues_review}
\begin{equation}
\label{eq:activestress}
\mathbb{\sigma}^{\rm active}=-\zeta \mathbb{Q}
\end{equation}
with an activity coefficient $\zeta$.  In principle, an isotropic
contribution to the stress can also exist, but it can be shown that
this contribution plays a subdominant role and can be ignored for
simplicity (see Appendix B).
The ribbon can be described by a quasi 2D sheet, whose shape can be described (in a Monge representation) by a function $h(x,y)$ for the vertical deflection relative to the center of the channel, and a sheet stiffness $\kappa_{\rm eff}$ that can be derived from the bending modulus of the individual filaments and the nematic order parameter (see Appendix B).\\
Focusing for simplicity on the variations along the $x$-axis, we can
construct a generalized free energy functional as
\begin{equation}
\label{eq:f}
\mathcal{F}[h]= \int \dd A \left[\frac 12\kappa_{\rm eff} (\partial ^2 _x h)^2 +\frac 12 \sigma^{\rm active}_{xx} (\partial_x h)^2\right]
\end{equation}
where the coupling between the active stress and the deformation arises from the nonlinear contribution to the strain tensor (see Appendix B for a full derivation that includes deformation in both directions). 
The emerging instability of the sheet is subject to viscous dissipation (either in the fluid, or in the remaining microtubule network outside of the sheet). We describe it with a local drag term that resists vertical movement of the sheet, which is written as $\Gamma \partial_t h$. In Appendix B (Role of hydrodynamic drag) we show \textit{a posteriori} that it largely exceeds the expected hydrodynamic drag in the fluid layer surrounding the sheet and we therefore attribute it to the residual filament network filling the volume. Within a linear stability analysis, we use the {\it ansatz} $h(x,t)\propto e^{\Lambda t} e^{i q x}$ for a perturbation with the wave vector $q$. We find a growth rate of
\begin{equation}
\label{eq:growth}
\Lambda(q)=\frac 1 \Gamma \left(-\kappa_{\rm eff} q^4 +\frac 12 \zeta S q^2 \right)\;.
\end{equation}
For negative values of the active stress corresponding to extensile stress, the growth rate $\Lambda(q)$ will be positive for an interval of wave
numbers, with the fastest growing mode having the wave number $q^*=\sqrt{\frac{\zeta S}{4 \kappa_{\rm eff}}}$ and a
growth rate
\begin{equation}
\label{eq:lambda2}
\Lambda(q^*)= \frac{(\zeta S)^2}{16 \Gamma \kappa_{\rm eff}}\;.
\end{equation}
Assuming that the samples do not differ in other parameters, we obtain a relationship between the characteristic time of pattern formation and the wavelength $\lambda=2 \pi/q^*$ as follows
\begin{equation}
\label{eq:taulambda}
\tau_{\rm w} \propto \lambda^4
\end{equation}
The experimental data shown in Fig.~\ref{Fig:Fig4}D confirm the trend predicted by Eq.~(\ref{eq:taulambda}), both when comparing the outcome in equal samples and in those with different ATP concentrations. With the above equations we can quantitatively estimate the active stress per filament (see Appendix B, Net force per filament) as $0.004\,\rm pN$, several orders of magnitude below the force of a single kinesin motor. This indicates that the motor force almost cancel out in the network and only a small imbalance contributes to the macroscopic stress.\\ 
Our results also suggest that the nematic order is the requirement for wrinkling and the instability is independent from the size of the channel. This explains the coexistence of buckling along perpendicular axes in adjacent areas of the same channel shown in Fig.~\ref{Fig:Fig4}F. 
In this case their nematic domains are oriented along perpendicular directions.\\

\paragraph*{\textbf{Simulation.}} We used a complementary computer simulation to directly verify that sheet formation and wrinkling arise from the interplay at nanometer scale between the action of kinesin motors and attractive depletion forces. We   simulated the dynamics of 20,000 filaments in a sufficiently large box to reproduce the salient phenomena. 
\begin{figure*}
	\centering
	\includegraphics[width=\linewidth]{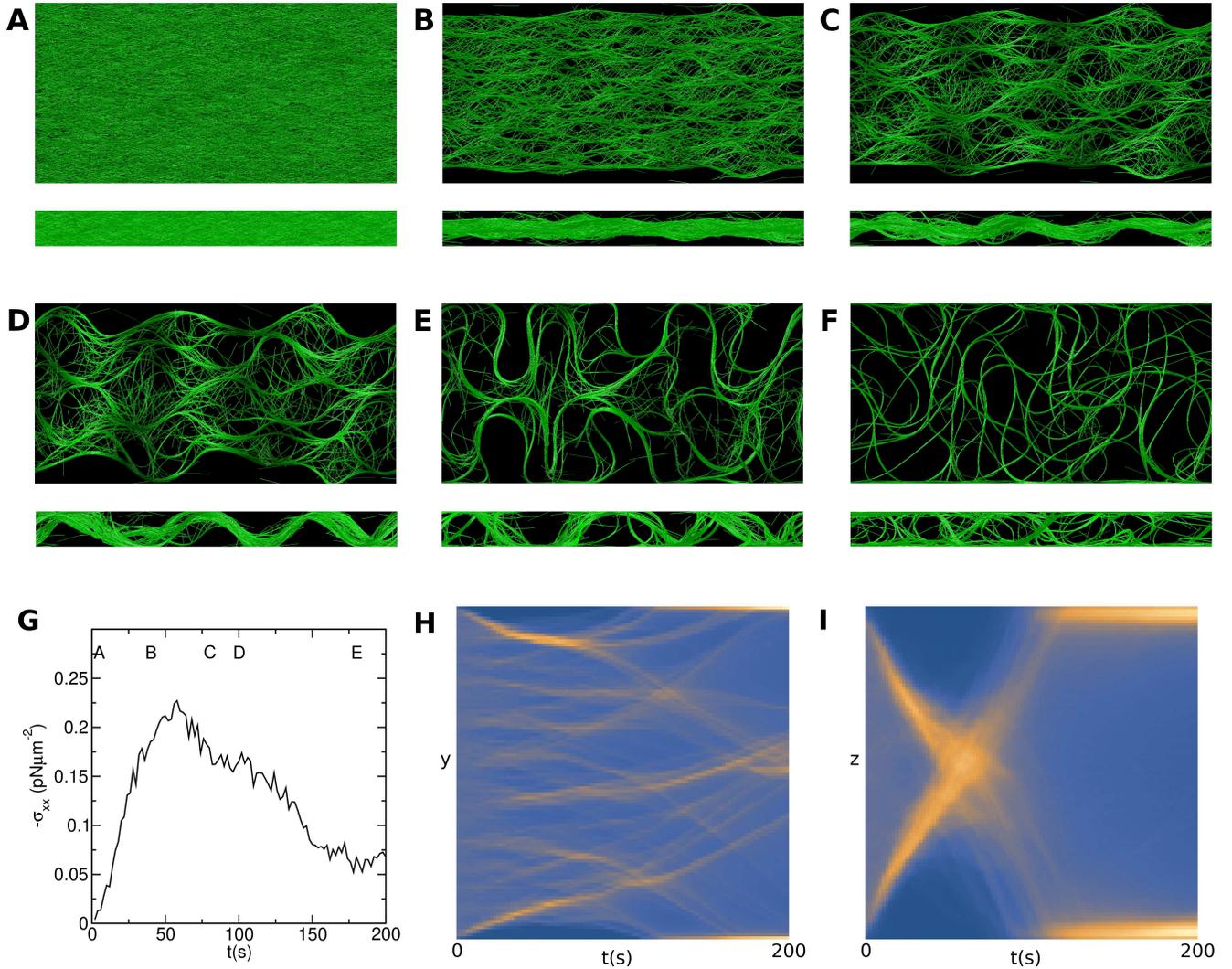}
	\caption{\setlength{\baselineskip}{0pt} \textbf{Simulation of the active nematic using the Cytosim package}. It reproduces the dynamics observed in the experiment. (\textbf{A}) Initially, the 3D volume is filled with nematically ordered filaments (with orientations within $\theta_{\max}=30^\circ$ of the $x$-axis), mixed with tetrameric active motors and passive cross-linkers. Filaments are subject to repulsive hard-core interactions and attractive depletion forces. (\textbf{B}) The filaments first form a ribbon that contracts laterally. (\textbf{C}) Due to extensile stress, the ribbon forms wrinkles. At the same time, increased bundling of filaments is visible. (\textbf{D}) The wrinkles grow until they get constrained by the top and bottom walls of the channel. (\textbf{E}) Under constraint, the bundles increasingly extend in horizontal direction and the ribbon structure is dissolved. (\textbf{F}) Finally, a state of 3D active turbulence is reached. Upper row: top view; lower row: side view. (\textbf{G}) Average stress $-\sigma_{xx}$ in a cross-section of the channel as a function of simulation time. Positive values represent extensile stress. The letters indicate the times of snapshots in panels (A-E). (\textbf{H-I}) Kymographs showing the average filament density across the width (H) and height (I) of the channel as a function of time. Contraction in both directions is maximal at the onset of the wrinkling instability. }
	\label{Fig:Fig5}
\end{figure*}
The simulated system consists of elastic filaments (microtubules), tetrameric kinesin motors (pairs of kinesin dimers that can bind to two adjacent filaments) and passive cross-linkers. Filaments are subject to hard-core repulsion and attractive depletion forces. The simulations use a Langevin-dynamics algorithm from the Cytosim package. The results are shown in Fig.~5 and Figure S6.\\
Starting from a nematically ordered state (Fig.~\ref{Fig:Fig5}A), the simulation first shows the condensation in the middle of the channel and shrinking in the lateral direction (Fig.~\ref{Fig:Fig5}B). As in the experiment, formation of microtubule bundles also becomes visible at this stage. Afterwards, the extensile stress (Fig.~\ref{Fig:Fig5}G) leads to a wrinkling instability in the sheet (Fig.~\ref{Fig:Fig5}C). The growth of the selected out-of-plane deformation mode continues until the amplitude becomes limited by the simulation box size (Fig.~\ref{Fig:Fig5}D). Afterwards, the ribbon structure disintegrates and the system ends in a state of 3D active turbulence (Fig.~\ref{Fig:Fig5}E-F). The simulation shows a very good qualitative agreement with the experimental observations. The lateral and vertical distribution of filaments as a function of time are shown as kymographs in Figs.~\ref{Fig:Fig5}H and I, respectively. The maximal contraction 
is 73\% vertically and 19\% horizontally. The contraction is not uniform, but proceeds from the borders towards the center, explaining why the horizontal contraction is relatively weaker, although the absolute displacement is larger.\\
The simulations are in agreement with our earlier conclusion that the kinesin motors responsible for stress generation are close to the stall conditions. At the onset of instability, the kinesins connecting antiparallel filaments act with an average force of around 4pN or 80\% of the stall force (Fig.~S6B). Furthermore, the simulation provides an insight into the mechanism behind the buildup of extensile stress. Because motors move towards the microtubule plus ends, a density gradient is established along the microtubules (Fig.~S6C). This, in turn, causes on average a compressive force in the microtubule, or an extensile stress in the network.\\
We also repeated the simulations in the absence of motor proteins (Fig.~S6E) and without the attractive interaction caused by PEG (Fig.~S6F). Without motors, bundles are still formed and the network contracts, albeit slower than with motors. Without attractive forces, the contraction is further slowed down and no bundling is visible.

\section*{Conclusion}
We present a 3D active system without substrate interactions that collapses into a sheet due to activity and further evolves to exhibit a 3D wrinkling instability before transitioning to fully developed active turbulence. 
The active stress is determined by the stall force of the molecular motors and simultaneous cross-linking function as the simulation confirmed. We expect our observation and explanation of these novel emergent properties to broaden the perspective of active nematic systems in 3D.

\section{Materials and Methods}
\textit{\textbf{Motile Bundle Solution.}}
The motile bundle solution was prepared as described before \cite{SANCHEZ2013205}. 
Briefly, the kinesin-streptavidin complexes 
were mixed with 2.7 mg/ml tubulin and 0.6\% PEG. (See Supplementary Methods for a more detailed description). 
By adding PEG, attractive interactions between microtubules are induced through depletion force and lead to bundle formation \cite{Asakura, Needlemann}. 
After the polymerization of microtubules, the active solution was injected into a PDMS microfluidic channel with a rectangular cross-section. 
To reduce unspecific protein adsorption the channel was functionalized with PLL-g-PEG (see Supplementary Methods).
The peak-to-peak amplitude of the wrinkling can be determined by vertical scans as the difference between the two z-positions at which the structure is in focus (Fig.~\ref{Fig:Figure1}F-G). 
\newline
\textit{\textbf{Imaging and tracking.}}
Image acquisition was performed using an inverted fluorescence microscope Olympus IX-71 with a 4$\times$ objective (Olympus, Japan) and a DeltaVision imaging system (GE Healthcare). 
The images were acquired with a variable frame rate according to the experiment with an exposure time of 500 ms for a variable time according to the experiment (See Supplementary Methods for more details).\\
\textit{\textbf{Simulation.}}
We simulated the evolution of a 3D active nematic system using the
open source Cytosim package \cite{Nedelec.Foethke2007}
(www.cytosim.org) (See Supplementary Methods for more details). 

\setcounter{equation}{0}
\renewcommand{\theequation}{S\arabic{equation}}
\setcounter{figure}{0}
\renewcommand{\thefigure}{S\arabic{figure}}
\setcounter{table}{0}
\renewcommand{\thetable}{S\arabic{table}}

\section{APPENDIX A}

\section{Supplementary methods}    
\paragraph*{\textbf{Microfluidic device.}}
The experimental chamber was constituted by a custom-made microfluidic device made of polydimethylsiloxane (PDMS, 10:1 mixture with curing agent, Sylgard 184, Dow Corning Europe SA). Standard soft lithography was used to produce the microfluidic channel that was 1.5 mm wide, 100 $\mu$m high, and 30 mm long.  Inlets and outlets for the active mixture were punched through the PDMS by using a syringe tip.

\paragraph*{\textbf{Non-adsorbing surface coatings and experimental chamber assembly.} }Glass coverslips ($64 \times 22\, \rm mm^2$, VWR) were cleaned by sonication in a 2\,$\%$ Hellmanex III solution (Hellma Analytics) for 30 minutes. Afterwards, they were extensive washed in deionized water, incubated 10 minutes in acetone, 10 minutes in ethanol, extensive washed in deionized water and drying with a filtered airflow.  The cleaned coverslips were immediately activated in oxygen plasma (FEMTO, Diener Electronics, Germany) for 30 s at 0.5 mbar and sealed to the PDMS. 0.1 mg/mL Poly(L-lysine)-graft-poly(ethylene glycol) (PLL-g-PEG) (SuSoS AG, Switzerland) in 10 mM HEPES, pH 7.4, at room temperature was injected into the channel and incubated for 1 h. Finally, M2B buffer was used to remove the excess PLL-g-PEG from the channel.

\paragraph*{\textbf{Motile bundle solution.}}
The motile bundle solution was prepared as described before \cite{SANCHEZ2013205}. 
Kinesin 401 was purified as previously published \cite{Gilbert, Young} and the kinesin-streptavidin complexes were prepared by mixing 0.2 mg/ml kinesin 401, 0.9 mM dithiothreitol (DTT), 0.1 mg/ml streptavidin (Invitrogen, S-888) (stochiometric ratio kinesin:streptavidin 2:1)  dissolved in M2B (80 mM PIPES, adjusted to pH = 6.9 with KOH, 1 mM EGTA, 2 mM MgCl\textsubscript{2})  and incubated on ice for 15 min. The plasmid that codes biotin-labeled kinesin 401 (Kinesin 401-BIO-6xHIS) was a gift from Jeff Gelles (pWC2 - Addgene plasmid \# 15960; http://n2t.net/addgene:15960; RRID\textunderscore Addgene\textunderscore15960)  \cite{Subramanian445} and was purified at the Dortmund protein facility (DPF).

The active mixture (AM) was obtained by mixing 2.4 mM Trolox (Sigma 238813), 1.7 $\mu$l pyruvate kinase/lactic dehydrogenase (PK/LDH, Sigma, P-0294), 32 mM phosphoenol pyruvate (PEP, VWR AAB20358-06), 16.6 $\mu$l 3$\%$ PEG, 5.5$\mu$l M2B and 3.25 $\mu$l DTT (10 mM), 0.5 mg/ml glucose, 0.2 mg/ml glucose oxidase (Sigma G2133), 0.05 mg/ml catalase (Sigma C40), 2 mM ATP and 4$\mu$l kinesin1-streptavidin clusters. The microtubule mixture (MT) was prepared by mixing 2.7 mg/ml 488 HiLyte\textsuperscript{TM} labeled porcine brain tubulin (Cytoskeleton, Inc., U.S.A.) in M2B with 5 mM MgCl\textsubscript{2}, 1 mM GTP, 50 $\%$ DMSO and 0.3$\%$ PEG.
The final mixture consists of 15$\mu$l MT, 29$\mu$l AM and 5.8 $\mu$l of a stabilising mixture composed by 68 mM MgCl$_2$ and 5 $\mu$M taxol. This solution is kept in the oven for 30 minutes by 37$^{\circ}$C and introduced into the PDMS channel thereafter.

Unlike previous experiments with 2D active nematics \cite{Sanchez} that used short microtubules polymerized with GMPCPP (Guanosine-5'-[($\alpha,\beta$)-methyleno]triphosphate, a GTP analogue), we used longer microtubules with an average length of 19 $\mu$m $\pm$  10 $\mu$m, polymerized with GTP and stabilized with taxol (see Fig.~S2). The injection was accomplished at low pressure by using a 50 $\mu$l syringe in order to avoid shear damage to the filaments. 

\paragraph*{\textbf{Imaging and tracking.}}
Image acquisition was performed using an inverted fluorescence microscope Olympus IX-71 with a 4$\times$ objective (Olympus, Japan) and a DeltaVision imaging system (GE Healthcare). For excitation, a Lumen 200 metal arc lamp (Prior Scientific Instruments, U.S.A.) was applied. The data was recorded with a CCD camera (CoolSnap HQ2, Photometrics). The images were acquired with a variable frame rate according to the experiment with an exposure time of 500 ms for a variable time according to the experiment.
The wavelengths of the pattern were measured by using a purpose-made algorithm written in python\textsuperscript{TM}. The images acquired during the experiments have been filtered with a threshold and binarized. The algorithm identifies the position of the intensity maxima corresponding to the waves. The wavelength is defined as the distance between such intensity peaks (see Fig. S1 below). The procedure is repeated at least five times for different $y$ positions within each the sample. The wavelength values are plotted as mean $\pm$ SD for each sample. Each data point in the plot in Fig.~3D correspond to an experiment.

\paragraph*{\textbf{Simulation.}} In the simulation we kept a filament density close
to the experimental value, but chose a smaller box which was
sufficiently large to capture the wrinkles while keeping the demand
for memory and CPU time feasible. We thus simulated a box of
200 $\mu$m length, 100 $\mu$m width and 20 $\mu$m
height with periodic boundary conditions in $x$ direction and
repulsive boundaries in $y$ and $z$ directions. The box contained
20,000 microtubules and the same number of tetrameric kinesin motors
and passive cross-linkers. Based on the experiment, the filament length
was chosen as 15 $\mu$m. The filaments are subject to repulsive
forces when the distance between their centerlines is closer than
$d_0=50\,\rm nm$. Beyond that distance, there is an attractive
interaction up to a distance $d_0+r_{\rm att}=90\,\rm nm$, which
describes the depletion forces caused by the PEG solution. Kinesin
motors attach to one or two microtubules with the rate
$r_{\rm on}=5\,\rm s^{-1}$ when they are within a range of
$100\,\rm nm$. When bound, they move along each filament with a
prescribed linear force-velocity relationship. The detachment from
microtubules is stochastic with a constant unbinding rate
$r_{\rm off}=0.1\,\rm s^{-1}$. Passive cross-linkers behave in a
similar way as the motors, but do not move along microtubules. The
simulation ran in steps of $\Delta t=0.02\,\rm s$ until the formation
of wrinkles. The simulation parameters are listed in Table S1.

\paragraph*{\textbf{Reconstruction of the nematic director field.}}
We used correlation analysis to quantify the nematic order in the sample. The image is given by the intensity function $I(X,Y)$ on a discrete lattice. To assess the order at position $(X_0,Y_0)$, we first evaluated the correlation function $C(\Delta x, \Delta y)$ as
\begin{equation*}
	\begin{split}
		C(\Delta x, \Delta y)= \frac 1 {(2a+1)^2} \sum_{k,l=-a}^a  I(X_0-\frac{\Delta x}{2}+k,Y_0-\frac{\Delta y}{2}+l)\\ \times I(X_0+\frac{\Delta x}{2} +k,Y_0+\frac{\Delta y}{2} +l) \\ - \frac 1 {(2a+1)^4}\left[  \sum_{k,l=-a}^a I(X_0-\frac{\Delta x}{2}+k, Y_0-\frac{\Delta y}{2} +l)  \right] \\ \times  \left[  \sum_{k,l=-a}^a  I(X_0+\frac{\Delta x}{2}+k,Y_0+\frac{\Delta y}{2} + l)  \right] \;,
	\end{split}
\end{equation*}
with the field size $a=16$.  Next, we fitted the correlation function
to a Gaussian
\begin{equation*}
  \begin{split}
    C_{\rm fit}(\Delta x, \Delta y)=C_0 \exp\bigl[ -(\Delta x \cos \phi +\Delta y \sin \phi)^2 / (2 \sigma_{\rm major}^2) \\ - (\Delta x \sin \phi -\Delta y \cos \phi)^2 / (2 \sigma_{\rm minor}^2) \bigr]\;,
  \end{split}
\end{equation*}
with the fit parameters $C_0$ (amplitude), $\phi$ (orientation of the major axis), $\sigma_{\rm major}$ (major correlation length), and $\sigma_{\rm minor}$ (minor correlation length, with  $\sigma_{\rm minor}<\sigma_{\rm major}$). We determined the nematic director as $\hat n=(\cos\phi,\sin\phi)$  and the order parameter as
\begin{equation*}
	S=1- \frac{\sigma_{\rm minor}}{\sigma_{\rm major}}\;.
\end{equation*}

\section{APPENDIX B}

\section{Stability analysis of the active elastic sheet}

In the following we present a more general discussion of  the dynamics of an active nematic sheet. The calculation takes into account the active stress, anisotropic in-plane elasticity and anisotropic bending stiffness of the sheet. 

We consider an elastic sheet that spans the $x-y$ plane before deformation. We describe in-plane deformations with the functions $u_1(x,y)$ and $u_2(x,y)$ and the out-of-plane deformation with the function $h(x,y)$. To the leading order, the strain in the deformed sheet is given by
\begin{equation}
\label{eq:1}
u_{ij}=\frac 12 \left[ \partial_i u_j +\partial_j u_i +(\partial_i h)(\partial_j h)\right]\;,
\end{equation}
and consists of a linear term for in-plane deformations $u_{ij}^{\ell} = \frac 12 (\partial_i u_j+\partial_j u_i)$ and a quadratic term $\frac 12 (\partial_i h)(\partial_j h)$ describing the strain caused by small out-of-plane deformations. We have ignored a terms in the form of $\frac12 (\partial_i u_k)(\partial_j u_k)$.

The curvature tensor is defined by 
\begin{equation}
\label{eq:2}
H_{ij}=\frac{\partial_i \partial_j h}{\sqrt{1+(\nabla h)^2}}= \partial_i \partial_j h+{\cal O}(h^3),
\end{equation}
and determines the mean curvature $\frac 12 H_{ii}=\frac 12 \Delta h$ and the Gaussian curvature $K=\det H$. Note the use of summation convention over identical indices. 

In the case of an isotropic sheet, the stretching elastic energy as given by the Hooke's law reads
\begin{equation}
\label{eq:3}
\mathcal{F}_{\rm stretch}^{I}=\int \dd A \left[ \frac 12 \lambda u_{ii} u_{jj} + \mu u_{ij}u_{ij}\right]\;,
\end{equation}
where $\lambda$ and $\mu$ are Lam\'e's first and second parameter, respectively. This elastic free energy corresponds to a linear relationship between the strain tensor $u_{ij}$, and the stress tensor
\begin{equation}
\label{eq:3a}
\sigma^I_{ij}= C_{ijkl}^I u_{kl}\;,
\end{equation}
where
\begin{equation}
\label{eq:3b}
C_{ijkl}^I=\lambda \delta_{ij} \delta_{kl}+\mu(\delta_{ik}\delta_{jl}+\delta_{il}\delta_{jk})
\end{equation}
is the isotropic 4-th order stiffness tensor. 
Using the homogeneity of Hooke's law, the stretching free energy can be written in terms of the stress tensor $\sigma_{ij}^I$ as 
\begin{equation}
\label{eq:3c}
\mathcal{F}_{\rm stretch}^{I}=\frac12 \int \dd A \, \sigma^I_{ij} u_{ij}\;.
\end{equation}
Note that this result is valid despite the fact that the expression for the strain contains nonlinear contributions in terms of the deformation field.

We can now generalize this framework to include anisotropy and activity. We describe the nematic order with the symmetric tensor $\mathbb{Q}=(\hat n \otimes \hat n - \frac 1 d \mathbb{I}) S$, with $\hat n$ denoting the director. $S$ is the order parameter, which is $1$ if all filaments are aligned in the direction $\hat n$. We introduce the anisotropic stiffness tensor $C_{ijkl}$, which, by definition, has the symmetries  $C_{ijkl}= C_{klij}= C_{jikl}= C_{ijlk}$. It also has to share the symmetries of the $\mathbb{Q}$-tensor. We therefore use the following ansatz:
\begin{widetext}
\begin{equation}
\label{eq:4}
C_{ijkl}=\chi Q_{ij}Q_{kl} +\gamma (Q_{ij}\delta_{kl}+\delta_{ij} Q_{kl}) +\lambda \delta_{ij} \delta_{kl}+\mu(\delta_{ik}\delta_{jl}+\delta_{il}\delta_{jk})\;.
\end{equation}
The anisotropic stretching elastic energy follows as
\begin{equation}
\label{eq:5}
\mathcal{F}_{\rm stretch}=\frac 12 \int \dd A \; C_{ijkl} u_{ij} u_{kl}\;,
\end{equation}
or, alternatively, as
\begin{equation}
\label{eq:5a}
\mathcal{F}_{\rm stretch}=\frac12 \int \dd A \, \sigma_{ij} u_{ij}\;.
\end{equation}
where
\begin{equation}
\label{eq:5b}
\sigma_{ij}= C_{ijkl} u_{kl}\;,
\end{equation}
represents the stress-strain relation for the anisotropic case. Note that Eqs. (\ref{eq:3}), (\ref{eq:3c}), (\ref{eq:5}), and (\ref{eq:5a}) include all possible nonlinear terms allowed by (the corresponding) rotational symmetry (of each case), upto and including terms of the order of ${\cal O}(u^\ell h^2)$ and ${\cal O}(h^4)$. 

In an analogous way, we write the bending energy of the isotropic sheet as 
\begin{equation}
\label{eq:6}
\mathcal{F}_{\rm bend}^I=\int \dd A \left[ \frac 12 \kappa H_{ii} H_{jj} + \bar \kappa K \right]\;,
\end{equation}
and generalize it to the anisotropic case as
\begin{equation}
\label{eq:7}
\mathcal{F}_{\rm bend}=\frac 12 \int \dd A \;\kappa_{ijkl} H_{ij} H_{kl}\;,
\end{equation}
with
\begin{equation}
\kappa_{ijkl}=\alpha Q_{ij}Q_{kl} +\nu (Q_{ij}\delta_{kl}+\delta_{ij} Q_{kl}) +(\kappa+\bar \kappa) \delta_{ij} \delta_{kl}-\frac{1}{2}\bar \kappa (\delta_{ik}\delta_{jl}+\delta_{il}\delta_{jk})\;.
\end{equation}

The active stress also follows the symmetry of the $\mathbb{Q}$-tensor, and we can write it as
\begin{equation}
\label{eq:9}
\sigma^{\rm active}_{ij}=-\zeta Q_{ij}-\zeta^I\delta_{ij}\;,
\end{equation}
where $\zeta$ and $\zeta^I$ are coefficients that represent non-equilibrium activity. A positive value of $\zeta$ implies that the anisotropic stress is extensile in the nematic direction; for a negative $\zeta$ it would be contractile. $\zeta^I$ describes an isotropic active stress that can exist in the system because there is no volume conservation for the active component in the solution. The active stress adds a contribution
\begin{equation}
\label{eq:10}
\mathcal{F}^{\rm active}= \int \dd A \;\sigma_{ij}^{\rm active} u_{ij}\;,
\end{equation}
to the total generalized free energy, which can be written explicitly as 
\begin{equation}
\label{eq:10q}
\mathcal{F}^{\rm active}=- \int \dd A(\zeta Q_{ij}+\zeta^I \delta_{ij}) \left[u_{ij}^{\ell} +\frac 1 2 (\partial_i h )(\partial_j h) \right]\;,
\end{equation}
upon inserting Eqs. (\ref{eq:1}) and (\ref{eq:9}).

We can now combine all the contributions to the generalized free energy that depend on the out-of-plane deformations
\begin{equation}
\mathcal{F}_{\perp}=\frac12 \int \dd A \left[\kappa_{ijkl} H_{ij} H_{kl} - (\zeta Q_{ij}+\zeta^I \delta_{ij})(\partial_i h )(\partial_j h) \right]+{\cal O}(h^3)+{\cal O}(u^\ell h^2) \;,\label{eq:20}
\end{equation}
and write the leading order contributions in the Fourier space as
\begin{equation}
\mathcal{F}_{\perp}=\frac12 \int \frac{\dd^2{\mathbf{q}}}{(2 \pi)^2}\left[\alpha (Q_{ij} q_i q_j)^2+\nu  Q_{ij} q_i q_j {\mathbf q}^2 + \kappa {\mathbf q}^4 -\zeta Q_{ij} q_i q_j -\zeta^I {\mathbf q}^2 \right] \left|h(\mathbf{q})\right|^2\;.\label{eq:21}
\end{equation}
Assuming that the sheet is subject to local viscous drag, we can write the rate of dissipation as
\begin{equation}
\mathcal{D}\equiv 2 \mathcal{R}=\int \dd A \;\Gamma (\partial_t h) ^2 =  \int \frac{\dd^2{\mathbf{q}}}{(2 \pi)^2} \;\Gamma \left| \partial_t h (\mathbf{q})\right|^2\;,
\end{equation}
where we have defined the Rayleighian dissipation function
$\mathcal{R}$. The generalized conservative force and the
generalized friction force together need to be zero,
$\delta \mathcal{F}/\delta h+ \delta \mathcal{R} / \delta {\dot h}=0$.
Then the dynamics of the mode with the wave vector $\mathbf{q}$ is
determined by
\begin{equation}
\partial_t h(\mathbf{q},t)= - \frac 1 \Gamma \left[ \alpha (Q_{ij} q_i q_j)^2+\nu  Q_{ij} q_i q_j {\mathbf q}^2 + \kappa {\mathbf q}^4 -\zeta Q_{ij} q_i q_j -\zeta^I {\mathbf q}^2 \right] h(\mathbf{q},t)\;,
\end{equation}
with the growth rate
\begin{equation}
\Lambda(\mathbf{q})=  - \frac 1 \Gamma \left[ \alpha (Q_{ij} q_i q_j)^2+\nu  Q_{ij} q_i q_j {\mathbf q}^2 + \kappa {\mathbf q}^4 \zeta Q_{ij} q_i q_j-\zeta^I {\mathbf q}^2 \right]\;.
\end{equation}
With the nematic order along $x$-axis ($\hat n= \hat e_x$), we have $\mathbb{Q}= \frac S 2 \left( \begin{array}{cc} 1 & 0 \\ 0 & -1 \end{array} \right)$ and
\begin{equation}
\Lambda(\mathbf{q})=  - \frac 1 \Gamma \left[ \frac{1}{4} \alpha S^2 (q_x^2-q_y^2)^2 +\frac{1}{2}\nu S (q_x^4-q_y^4)  + \kappa (q_x^2+q_y^2)^2  -\frac{1}{2}\zeta S (q_x^2-q_y^2) -\zeta^I (q_x^2+q_y^2) \right]\;.
\end{equation}
\end{widetext}
Because the last two terms are quadratic in $\mathbf{q}$, whereas all others are 4-th power, they will dominate at small $\mathbf{q}$ and it is always possible to find wave vectors with a positive growth rate, i.e., unstable modes. The growth rate is always maximal in $x$-direction. On that axis, it reads
\begin{equation}
\Lambda(q_x)=  - \frac 1 \Gamma \left[ \left(\frac{1}{4} \alpha S^2 +\frac{1}{2} \nu S  + \kappa\right) q_x^4    -\frac{1}{2}\left(\zeta S+2\zeta^I\right) q_x^2 \right]\;.
\end{equation}
Experimentally, we observe that the instability is present only when a sufficient initial degree of nematic alignment exists in the system. This suggests that the isotropic active stress plays a less dominant role as compared to the nematic active stress. Indeed, we expect the isotropic component to be contractile, and therefore contribute with the opposite sign (since $\zeta >0$ and $\zeta^I <0$) and oppose the formation of the instability. If a non-zero isotropic stress exists, then the instability requires a minimal threshold degree of nematic order given by $S \geq 2 |\zeta^I|/\zeta$. We can ignore the isotropic stress for a minimal representation of the theory, as it is done in the main text, where we have also used the shorthand $\kappa_{\rm eff} = \kappa +\frac 12 \nu S+\frac 14 \alpha S^2$.

We can also examine the prediction of the theoretical framework with regards to the stretching degrees of freedom. From the experimental observations, we expect to have a force balance in the direction perpendicular to the nematic alignment, namely the $y$-direction, since the deformation in this direction happens faster than the bending deformation. This entails 
\begin{equation}
u^{\ell}_{yy}=-\frac{\zeta S}{4\mu}+\frac{\zeta^I}{2(\lambda+\mu)}\;,\label{eq:30}
\end{equation}
to the lowest order. Therefore, we find that the anti-symmetric stress causes contraction orthogonal to the nematic direction (since $\zeta >0$ and $\zeta^I <0$), as observed experimentally. It is important to note that the contribution from Eq.~(\ref{eq:30}) does not modify the spectrum of the out-of-plane deformations given in Eq.\ (\ref{eq:21}), as the inclusion of the nonlinear terms will lead to an exact cancellation of any additional term the depends on $h$. In the $x$-direction, however, the deformation mode relaxes at the same time scale as the bending mode, and, therefore, the stationary force balance condition does not hold. To solve for the longitudinal deformation mode $u^\ell_{xx}$, we need to have information about the boundary condition of the system, and in particular, whether the stress is balanced due to longitudinal friction or contact with the boundaries. In the absence of such information, we can only speculate about different possible scenarios. If mechanical contact is established with the substrate, then we expect $u^\ell_{xx} \simeq 0$. In the absence of such contact, the friction force can balance the residual stress in the $x$-direction, leading to an extension that grows with time. While we cannot rule out this possibility, we have not observed experimentally extension amplitudes that are comparable or larger than the extent of lateral contraction. This implies that the assumption to ignore the third order term of the form $u^\ell_{xx} (\partial_x h )^2$ in Eq.\ (\ref{eq:20}) is justified.

\section{Quantitative aspects}
\subsection{Net force per filament}

The theory can be used to estimate the average stress per filament. If we assume a thin sheet where the only contribution to the bending elasticity is that of bending filaments, it becomes
\begin{equation}
\label{eq:kappa}
\kappa_{\text{eff}}=\rho L\cdot EI \left< n_x^2 \right> 
\end{equation}
where $\rho$ is the surface density filaments in the sheet, $L$ their length, $EI$ the bending modulus and $n_x$ the component of the filament orientation in the direction of bending ($n_x=1$ for nematic alignment). 

Through the equation $q^*=\sqrt{-\sigma_{xx}/2 \kappa_{\text{eff}}}$, we obtain the stress
\begin{equation}
\label{eq:stress}
-\sigma_{xx}=2\rho L\cdot EI \left< n_x^2 \right>  (2\pi/\lambda)^2
\end{equation}
At the same time, the stress can be expressed from the average force in a filament,
\begin{equation}
\label{eq:fbar}
-\sigma_{xx}=\rho L \left< n_x^2 \right> \bar f
\end{equation}
resulting in $\bar f=2EI (2\pi/\lambda)^2$.  With the values
$EI=0.4\times 10^{-23}\,\rm Nm^2$ and $\lambda=300\,\rm \mu m$, the
stress per filament is $0.004\,\rm pN$, several orders of magnitude
smaller than the forces exerted by kinesins. This result shows that
the motor and crosslinker forces largely cancel out in the network and
only a small bias towards extensile drives the instability.

The relationship can be tested in the simulation (due to the reduced system size the values are somewhat scaled down). With  $EI=2\times 10^{-23}\,\rm Nm^2$ and $\lambda=67\,\rm \mu m$, the estimated force per filament $0.35\,\rm pN$ is in agreement with simulation value at the onset of wrinkling instability, which is $0.31\,\rm pN$.

\subsection{Role of hydrodynamic drag}

The effective vertical drag coefficient per unit area, denoted $\Gamma$, can be estimated from Eq.~(4), which can be rewritten as
\begin{equation}
\Gamma=\frac{\kappa_{\text{eff}}}{\Lambda} \left( \frac{2\pi}{\lambda} \right)^4
\end{equation}
With the stiffnes from Eq.~(\ref{eq:kappa}), the density $\rho L=400\,\rm \mu m^{-1}$ and growth rate $\Lambda=1/(1000\,\rm s)$, we estimate $\Gamma=300,000\,\rm Pa (m/s)^{-1}$.

The fluid in the space between the sheet (assumed as impermeable) and the wall is governed by the continuity equation $\dot h (x,t)=(1/w) \partial x Q(x,t)$ and the flow rate $Q(x,t)=-H^3w/(12\eta)\partial_x p$. For a perturbation with a wavelength $\lambda$, the drag density is
\begin{equation}
\label{eq:drag}
2\frac{p}{\dot h}=24 \eta / H^3 (\lambda/(2\pi))^2= 1300\,\rm Pa (m/s)^{-1}
\end{equation}
with $H=35\,\rm \mu m$. We conclude that the estimated hydrodynamic drag is significantly smaller than the total drag. A possible explanation is that the filaments outside the sheet provide effective drag that resists vertical displacements of the sheet. 
	
\section*{Acknowledgements}
	E.B., I.G., R.G. acknowledge support from the MaxSynBio Consortium which is jointly funded by the Federal Ministry of Education and Research of Germany and the Max Planck Society. E.B. acknowledges support from the Volkswagen Stiftung (priority call “Life?”). A.V. acknowledges support from the Slovenian Research Agency (grant no. P1-0099).

\bibliography{references}

\onecolumngrid
\clearpage

\section{Supplementary figures}

\begin{figure*}[h!]
	\centering
	\includegraphics[width=0.8\linewidth]{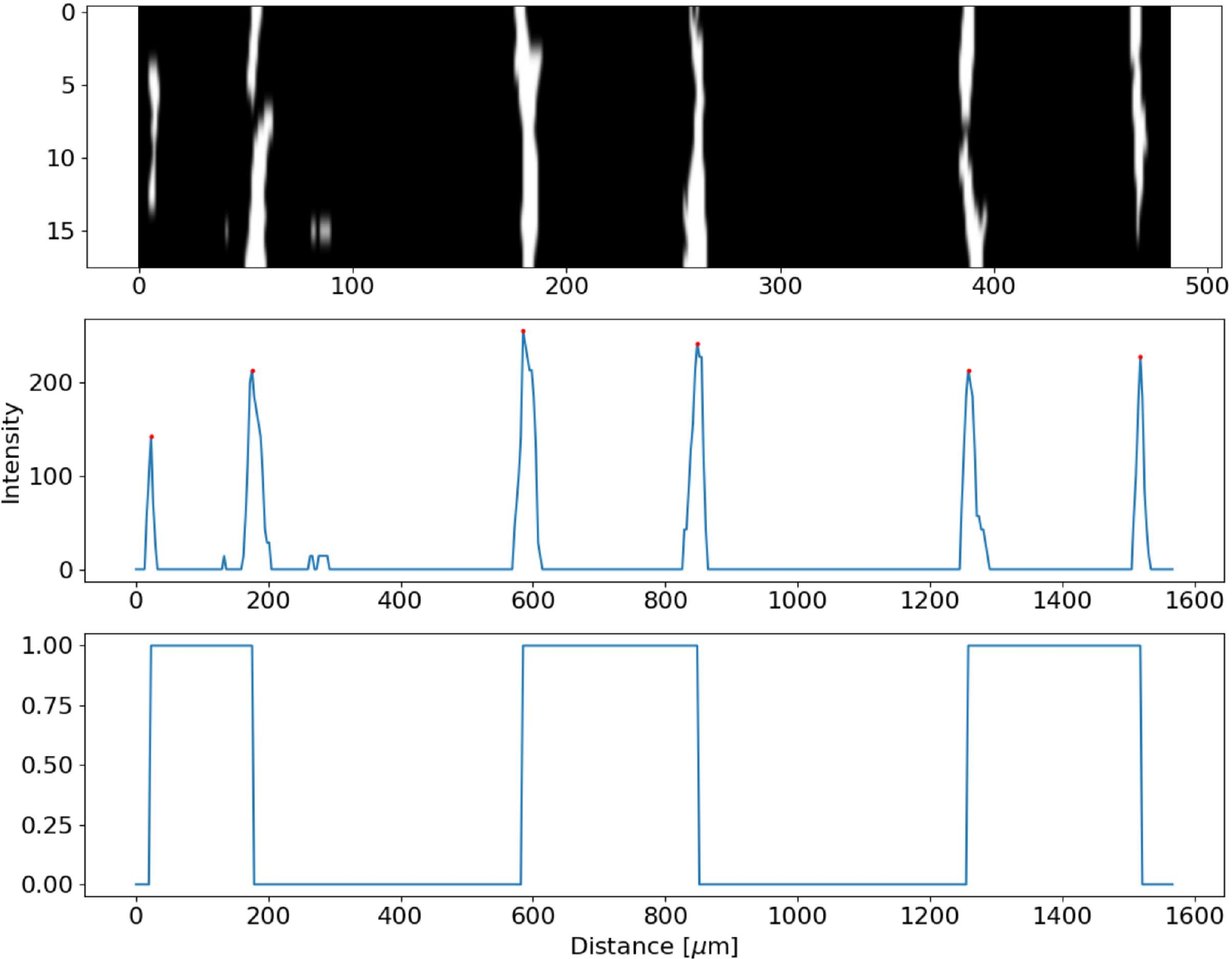}
	\caption{\textbf{Procedure for the evaluation of wavelengths}. The algorithm identifies the position of the intensity maxima corresponding to the waves. The wavelength is defined as the distance between the intensity peaks.}
\end{figure*}

\begin{figure*}[h!]
	\centering
	\includegraphics[width=0.5\linewidth]{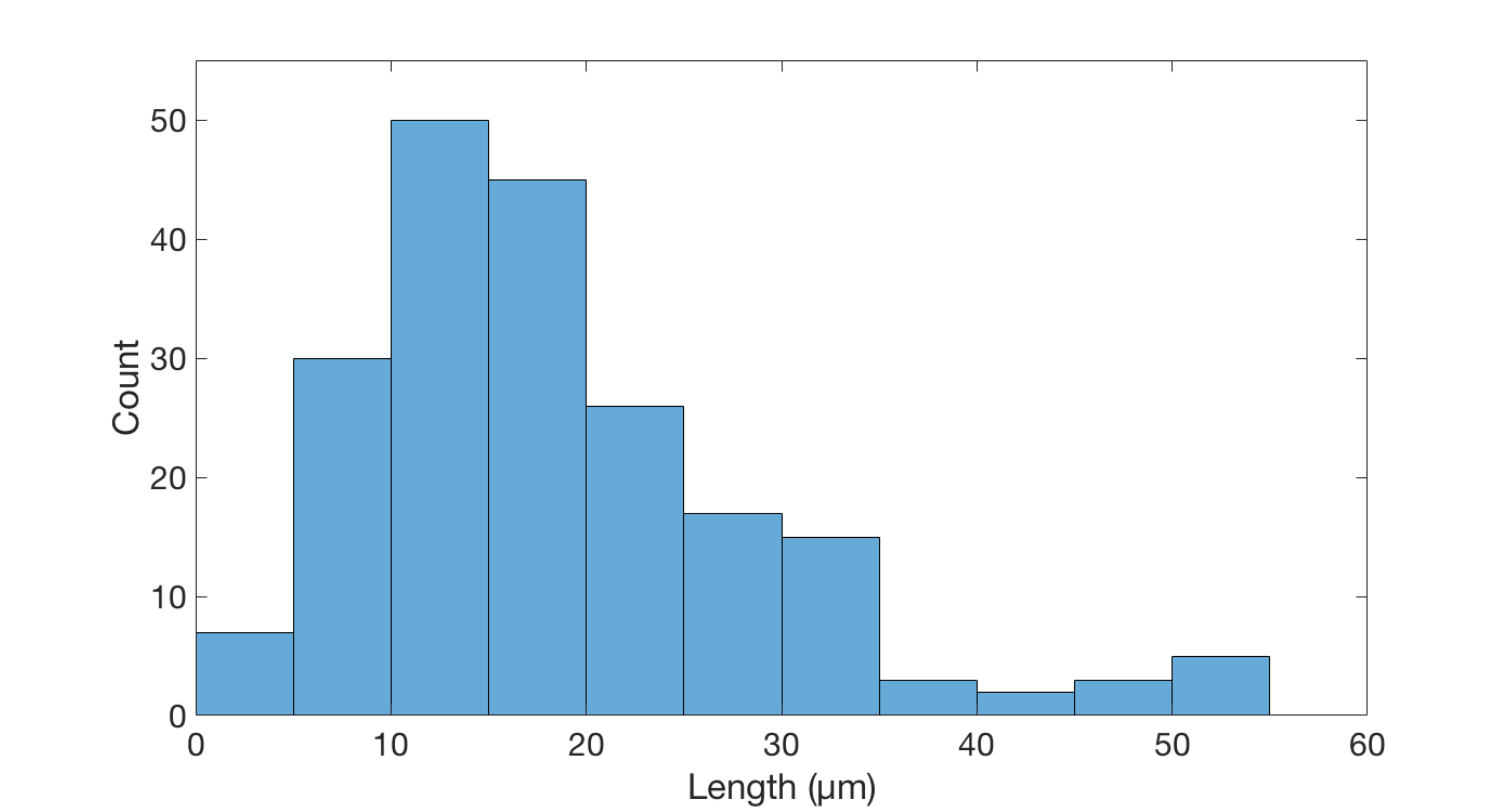}
	\caption{\textbf{Length distribution}. Microtubule length distribution after 30 minutes of polymerization.}
	\label{Fig:S1_MT_Length}
\end{figure*}

\begin{figure*}[h!]
	\centering
	\includegraphics[width=\linewidth]{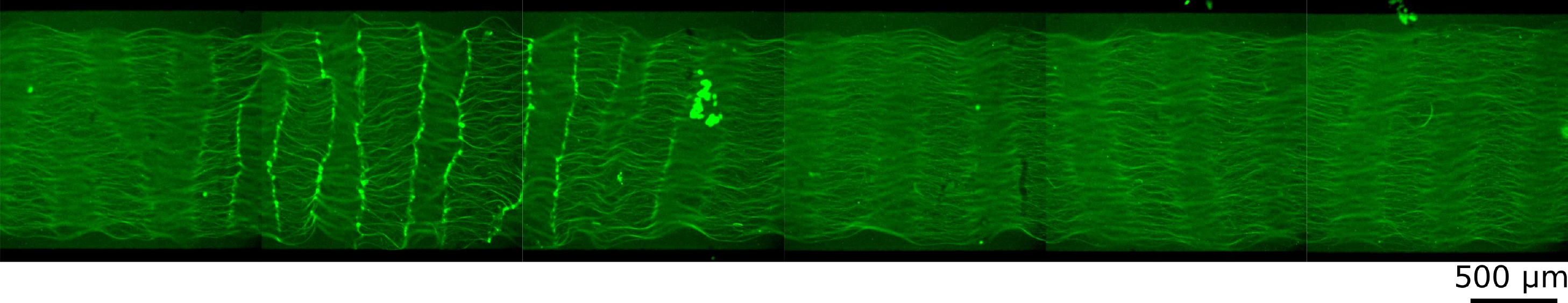}
	\caption{\textbf{Emergent pattern along the length of the channel}. The image is obtained by snapshots of adjacent areas acquired with an automated microscope stage.}
	\label{Fig:S2_panel}
\end{figure*}

\begin{figure*}[h!]
	\centering
	\includegraphics[width=0.5\linewidth]{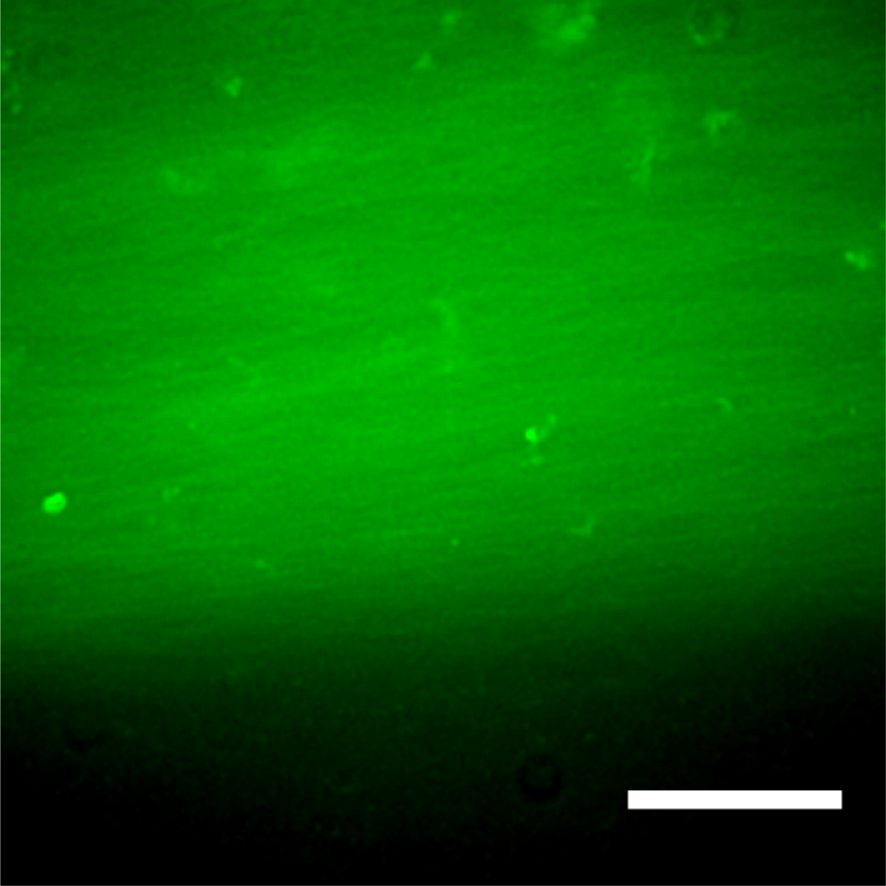}
	\caption{\textbf{Filament alignment without depletion effect}. Parallel arrangement of microtubules within the network at higher magnfication (60x). No bundles are visible as expected from the missing PEG. Scale bar: 25 $\mu$m. }
	\label{Fig:S3_woPEG}
\end{figure*}

\begin{figure*}[h!]
	\centering
	\includegraphics[width=\linewidth]{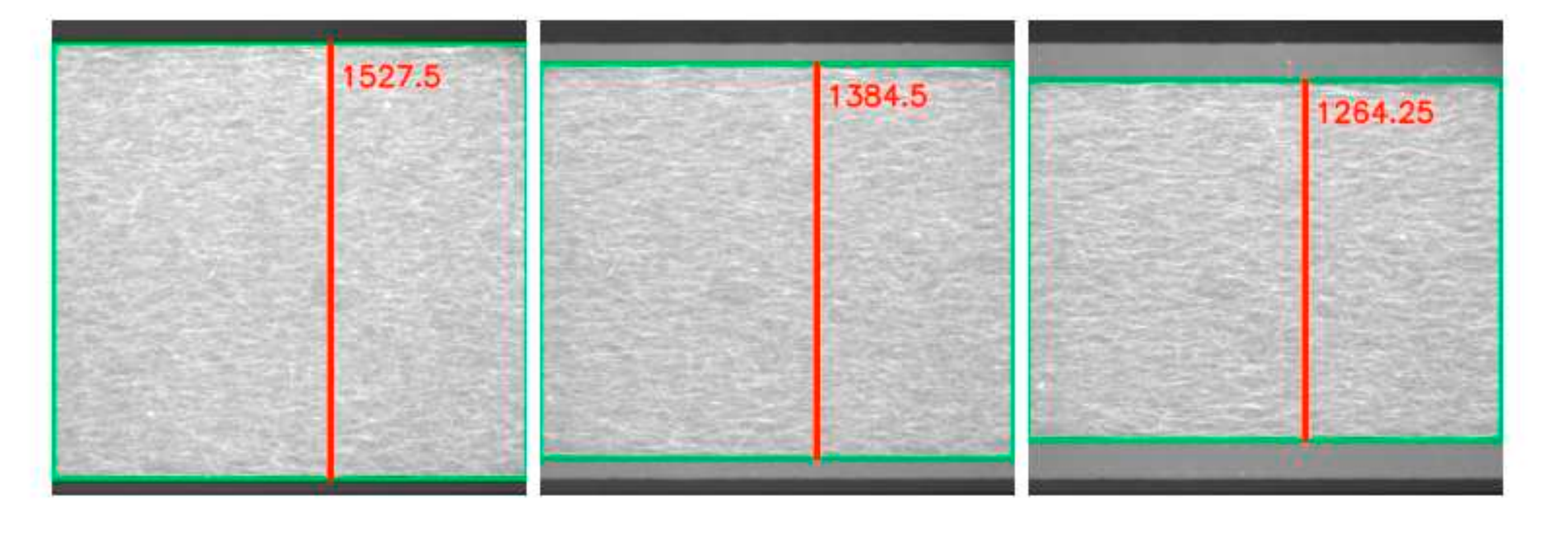}
	\caption{\textbf{Algorithm for contraction measurement}. Contraction measurement procedure based on the intensity value.}
	\label{Fig:S6}
\end{figure*}

\begin{figure*}[h!]
	\centering
	\includegraphics[width=\linewidth]{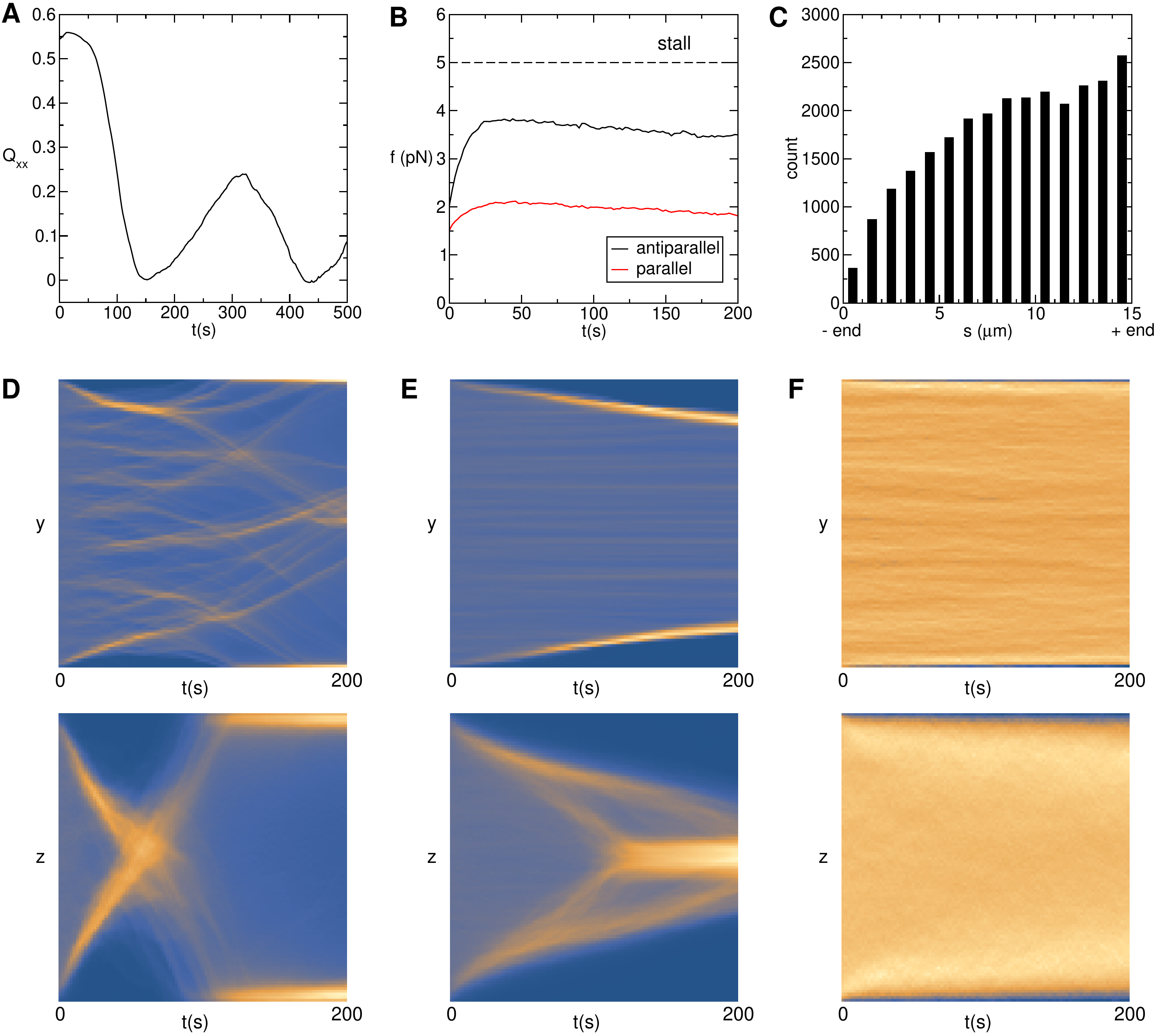}
	\caption{\textbf{Analysis of simulation results}. (A) Nematic order, characterized by the longitudinal component of the Q-tensor $Q_{xx}$ as a function of time. To exclude boundary effects, only filaments from the inner 80\% of the channel width and 80\% of the channel height were evaluated. (B) Average force generated by motors acting between parallel (red) and antiparallel (black) microtubules. The force between antiparallel filaments is close to the stall force (dashed) at the onset of the wrinkling instability. (C) Distribution of motor complexes along the length of microtubules. Only complexes that are simultaneously bound to two filaments are counted. The asymmetric distribution is the basis for teh emergence of extensile stress. (D-F) Kymographs showing the average filament density as a function of time across the width (top) and height (bottom) of the channel. (D) Reference conditions, as in Figure 5H,I. (E) Same conditions, but without motor proteins; the slow contraction is caused by depletion forces alone. (F) Simulation with motor proteins, but without depletion forces. Color scale varies between panels.}
	\label{Fig:S7}
\end{figure*}

\begin{table*}
	\caption{Simulation Parameters}
	\begin{center}
		\begin{tabular}{l l l p{\linewidth}}
			\hline
			Symbol & Parameter & Value & Notes\\ [0.5ex] 
			\hline \\
			$k_BT$ & Thermal energy & $4.2\, \rm pN\,nm$ & Room temperature \\ 
			
			$\eta$ & Viscosity & $0.01 \,\rm  pNs/ \mu m^{2}$ & Effective viscosity of the fluid \\
			
			$\Delta t$ & Time step & $0.02 \, \rm  s$ & Total time simulated $\sim 700\,\rm s$\\
			
			$b_x \times b_y \times b_z$ & Box size & $(200 \times 100 \times 20) \rm \mu m^3$ & Periodic b.c. in $x$-direction \\[3.5ex]
			
			\multicolumn{2}{l}{{\bf Filaments}} & \\[1.5ex]
			
			$N_{\rm MT}$ & Number of filaments & $2 \times 10^{4}$ & Estimated from the density\\
			
			$L$ & Filament length & $15 \, \rm  \mu m$ & Figure S1 \\
			
			$\theta_M$ & Initial nematic order & $ 30^\circ$ &  Uniform distribution of filament orientations on a sphere cap\\
			
			$EI$ & Filament stiffness & $20 \, \rm  pN \mu m^{2}$ & MT bending modulus  \\
			
			& Filament segmentation & $1 \, \rm  \mu m$ & Chosen \\
			
			$d_0$ & Filament diameter & $50 \, \rm  nm$ & Effective MT diameter \\
			
			$r_{\rm att}$ & Attraction range & $40 \, \rm  nm$ & Mimicking depletion force\\
			
			$k_{\rm att}$ & Strength of depletion force & $50 \, \rm  pN/ \mu m$ &\\
			
			$k_s$ & Strength of steric repulsion & $100 \, \rm  pN/ \mu m$ &\\[3.5ex]
			
			\multicolumn{2}{l}{{\bf Motors \& Crosslinkers}} & \\[1.5ex]
			
			$N_{\rm MP}$ & Number of kinesin complexes & $2 \times 10^{4}$ & \\
			
			$N_{\rm Cr}$ & Number of crosslinkers & $2 \times 10^{4}$ & \\
			
			$k_{\rm on}$ & Binding rate & $5 \, \rm  s^{-1}$ & \\ 
			
			& Binding range & $100 \, \rm  nm$ & \\
			
			$k_{\rm off}$ & Unbinding rate & $0.1 \, \rm  s^{-1}$ & \\
			
			$\upsilon_0$ & Unloaded speed of motor & $0.5 \, \rm  \mu m/s$ & Kinesin-1 speed at saturating ATP\\
			
			$f_s$ & Stall force & $5 \, \rm  pN$ & Kinesin-1 stall force\\
			
			$k$ & Link stiffness & $100 \, \rm  pN/ \mu m$ &\\
			$D$ & Diffusion coefficient & $10 \, \rm  \mu m^{2}/s$ & \\ 
			\hline
		\end{tabular}
	\end{center}
\end{table*}
\begin{table*}
	\caption{List of variables}
	\begin{center}
		\begin{tabular}{l p{\linewidth}}
			\hline
			Symbol & Parameter \\ [0.5ex] 
			\hline \\
			$(x,y,z)$ & Spatial coordinates along the length/width/height of the channel\\
			$h(x,y)$ & Vertical displacement of the sheet from the centerplane\\
			$H$ & Channel height\\
			$u_i(x,y)$ & In-plane displacement of the sheet from the centerplane\\
			$H_{ij}$ & Curvture tensor\\
			$K$ & Gaussian curvature\\
			${\cal F}_{\text{stretch}}$ & In-plane elastic contribution to the free energy\\
			${\cal F}_{\text{bend}}$ & Out-of-plane elastic contribution to the free energy\\
			${\cal F}^{\text{active}}$ & Contribution of active stress to the effective free energy\\
			${\cal F}_{\perp}$ & Out-of-plane contributions to the effective free energy\\
			${\cal D}$ & Dissipation rate \\
			$C_{ijkl}$ & In-plane stiffness tensor\\
			$\chi,\gamma,\lambda,\mu$ & Coefficients determining    $C_{ijkl}$ \\
			$\kappa_{ijkl}$ & Out-of-plane bending stiffness tensor\\
			$\alpha,\nu,\kappa,\bar\kappa$ & Coefficients determining $\kappa_{ijkl}$\\
			$\kappa_{\text{eff}}$ & Simplified bending stiffness\\
			$Q_{ij}$ & Q-tensor, describing the nematic order\\
			$S$ & Nematic order parameter\\
			$\zeta$, $\zeta^I$ & Activity coefficient (traceless and isotropic contribution, respectively)\\
			$\sigma_{ij}^{\text{active}}$ & Active stress in the sheet (unit force/length)\\
			$\Gamma$ & Drag coefficient per unit area for out-of-plane motion\\
			$\rho$ & Area density of microtubules in the sheet\\
			$L$ & Average microtubule length\\
			$\Lambda$ & Growth rate of the wrinkling mode (unit 1/time)\\
			$q$ & Wave vector\\
			$\lambda$ & Wavelength\\
			$\tau_C$, $\tau_w$, $\tau_{at}$ & Characteristic times for sheet contraction, wrinkling and active turbulence\\
		\end{tabular}
	\end{center}
\end{table*}

\end{document}